\documentclass[reprint,aps,nofootinbib]{revtex4-2}

\usepackage[
    top=1.8cm,
    bottom=2cm,
    left=1.5cm,
    right=1.5cm,
    paper=a4paper
    ]{geometry}

\usepackage[utf8]{inputenc}
\usepackage[english]{babel}
\usepackage[T1]{fontenc}
\usepackage[colorlinks=true, breaklinks=true, linkcolor=blue, citecolor=blue, urlcolor=blue]{hyperref}

\usepackage{xcolor}
\usepackage{graphicx}
\usepackage{amsfonts, amsmath, amssymb}

\usepackage{listings}
\newcommand{\lil}{\lstinline[basicstyle=\ttfamily]}

\newcommand{\be}[1]{\begin{equation}\label{#1}}
\newcommand{\ee}{\end{equation}}

\newcommand{\ket}[1]{| #1 \rangle}

\newcommand{\proj}[2]{| #1 \rangle \langle #2 | }

\begin{document}

\title{Cloud on-demand emulation of quantum dynamics with tensor networks}
\date{\today}

\author{Kemal Bidzhiev}
\email{kemal.bidzhiev@pasqal.com}
\author{Aleksander Wennersteen}
\email{aleksander.wennersteen@pasqal.com}
\author{Mourad Beji}
\author{Mario Dagrada}
\author{Mauro D'Arcangelo}
\author{Sebasti\'an Grijalva}
\author{Anne-Claire Le H\'enaff}
\author{Anton Quelle}
\author{Alvin Sashala Naik}

\affiliation{PASQAL, 7 rue Léonard de Vinci, 91300 Massy, France}

\begin{abstract}
We introduce a tensor network based emulator, simulating a programmable analog quantum processing unit (QPU). The software package is fully integrated in a cloud platform providing a common interface for executing jobs on a HPC cluster as well as dispatching them to a QPU device. We also present typical emulation use cases in the context of Neutral Atom Quantum Processors, such as evaluating the quality of a state preparation pulse sequence, and solving Maximum Independent Set problems by applying a parallel sweep over a set of input pulse parameter values, for systems composed of a large number of qubits.
\end{abstract}

\maketitle

\section{Introduction}


Quantum technology proposes leveraging the laws of quantum mechanics to process information in a framework that can enable solving hard computational problems more efficiently than classical alternatives. A new scientific age of quantum information has been precipitated by this idea, and constructing the first performant quantum processing units (QPUs) has become a key goal in recent years. As QPUs gain in effectiveness and reliability, experimental demonstrations have come forward solving specific computational tasks which may show an advantage over classical methods, such as sampling problems \cite{arute2019quantum, zhong2020quantum, wu2021strong, madsen2022quantum}, and quantum simulation implementations \cite{scholl2021quantum, ebadi2021quantum}. Platforms for QPUs have been proposed based on diverse physical architectures, including neutral atoms, superconducting circuits, trapped ions and photons.

As well as continuing the experimental development of QPUs, it is becoming increasingly important to simulate the behavior of particular physical QPU architectures, referred to as \emph{emulation}. Emulators encode realistic physical constraints including hardware-specific time dynamics and noise effects. Furthermore, they can inform about the expected performance of the QPU and benchmarking \cite{jaschke2022ab}. This allows the user to circumvent overhead by performing test jobs e.g. on a High-Performance Computing (HPC) backend before forwarding the full job on to the QPU. Moreover, integrating QPU and emulator backends through cloud access, offering the choice of backend to the user, allows efficient workflows as part of a full-stack quantum computer. The hybrid workflow would send suitable routines to the quantum processor, while executing larger computational tasks on the emulator backend. Systems supporting such hybrid workflows with emulated quantum devices already provide valuable services to researchers \cite{guerreschi2020intel, humble2021quantum, ravi2021quantum, mandra2021hybridq, mccaskey2018language}.

In most cases, the simulation of quantum systems has a complexity that grows exponentially with the system size. A significant instrument which has emerged to tackle this challenge are tensor networks~\cite{Schollwck2011}, providing a powerful structure to represent complex systems efficiently. The framework of tensor networks has allowed the efficient simulation and study of quantum systems, such as, most importantly for our purposes, the numerical study of time-evolution of quantum states which are ground states of gapped local Hamiltonians. Further, they have found uses in the identification of new phases of matter \cite{iregui2014probing, jiang2017anyon}, as a structure for processing data \cite{Cichocki2016, stoudenmire2016supervised}, or the numerical benchmarking of state-of-the-art QPU experiments with large qubit numbers \cite{scholl2021quantum, pan2022solving}. 

\begin{figure}[ht!]
    \centering
    \includegraphics[width=\linewidth,
                   trim={8.5cm 2cm 6cm 2.3cm},
                   clip]{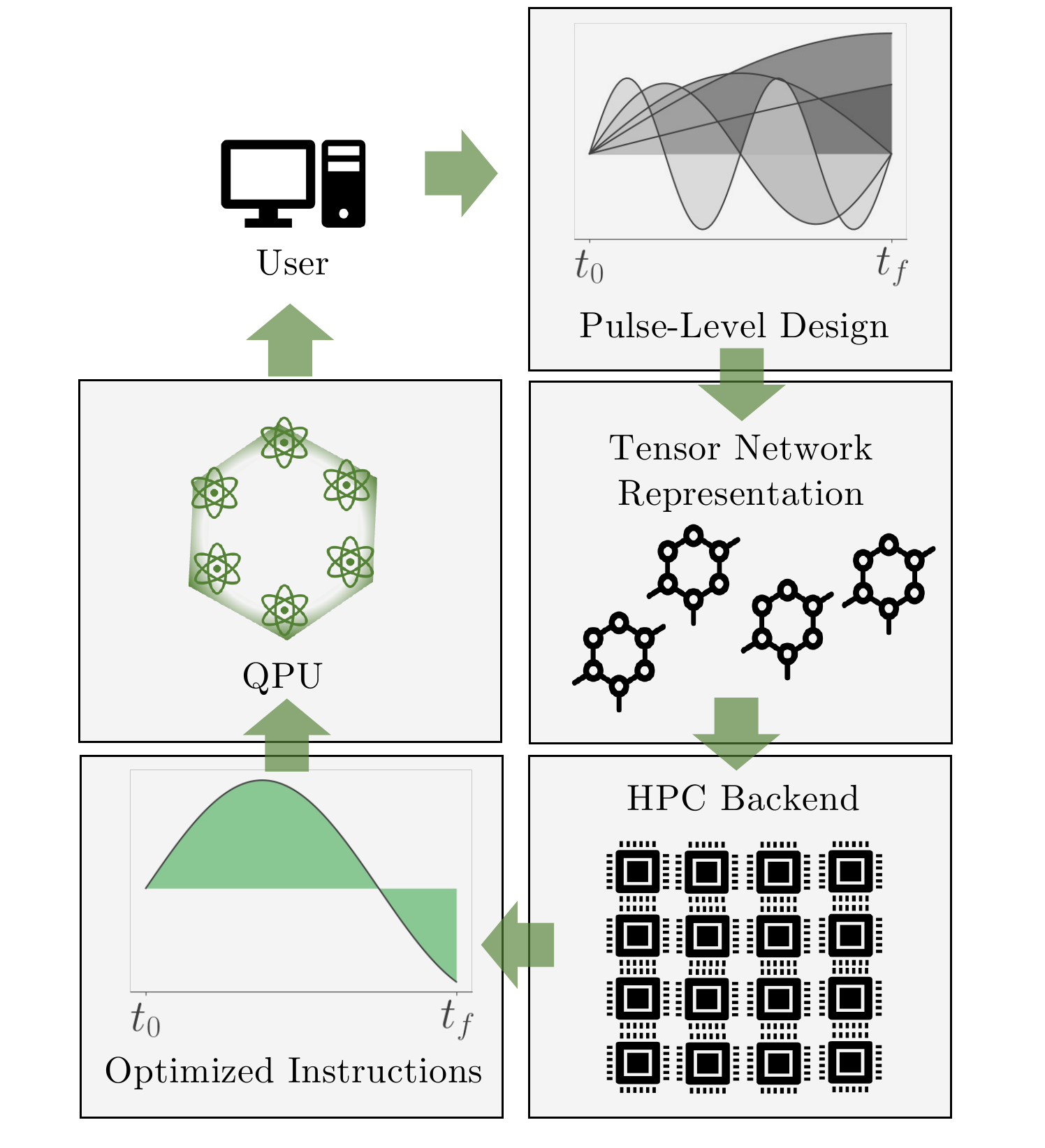}
    \caption{Role of emulation in a workflow for large quantum systems. Pulse-level design tools are complemented with tensor network representations of the corresponding quantum states. Algorithms performing time evolution take into account Hamiltonian dynamics as well as realistic hardware conditions, which are simulated on a HPC backend. These can be used to prepare optimized instructions on a QPU. We present a cloud-based platform integrating these tasks on-demand.}
    \label{fig:emu-tn-cycle}
\end{figure}

The paper is structured as follows. In Section \ref{sec:numerical-quantum} we review the elementary setup for the numerical simulation of quantum dynamics and introduce relevant tensor-network methods. Then, in Section \ref{sec:architecture} we describe the framework for integrating HPC-supported numerical algorithms into the cloud. We demonstrate applications of the emulator  in \ref{sec:applications}, simulating a neutral atom quantum processor in three examples: producing 1D $\mathbb{Z}_n$-ordered states, preparing 2D antiferromagnetic states and performing a parallel sweep of pulse parameters to search for Maximum Independent Sets on a particular graph. We conclude with comments about extensions to the system and a discussion about emulated environments in heterogeneous computing.

\section{Numerical simulation of Quantum Dynamics}\label{sec:numerical-quantum}

\subsection{Time-evolution of a Quantum System}\label{sec:hamiltonian-emulator}

\emph{Qubits}, or 2-level quantum systems, are the fundamental building blocks of the quantum systems that we will study. A \emph{quantum state} for an $N$-qubit system is an element of a $2^N$-dimensional Hilbert space. It can be represented by a complex vector $\ket{\psi} = \sum_x c_x \ket{x}$ of size $2^N$, where we choose $\{ \ket{x} \}$ to be an orthonormal basis of the space. We concentrate on an orthonormal basis that is built from the eigenstates of the $\hat \sigma^z$ operator on each site, which is called the \emph{computational basis}. Its elements can be written as $\ket{s_1 s_2 \cdots s_N}$, where each $s_i \in \{ 0, 1\}$. The strings $s_1 s_2 \cdots s_N$ are often called \emph{bitstrings}. A general quantum state is thus written as:
\be{eq:quantum_state}
|\psi \rangle = \sum_{s_1 \cdots s_N} c_{s_1 \cdots s_N} |s_1 \cdots s_N\rangle,
\ee
where the $c_{s_1\cdots s_N}$ is a complex number corresponding to a probability amplitude for a given set of indices $s_1 s_2 \cdots s_N$ representing the outcome of a measurement. The QPU takes an initial state and evolves it through a series of gates and/or analog pulses. To obtain the final state, it makes a projective measurement on it with respect to a certain measurement basis, repeating the cycle until a certain condition is met. Of these operations, the most computationally expensive calculation is to compute the final time-evolved state, which we describe next.

The time evolution of a quantum state is governed by the \emph{Schr\"odinger Equation}
\be{eq:schro}
i\frac{d \ket \psi }{d t} = \hat H(t) | \psi \rangle,
\ee
where we consider a particular time-dependent \emph{Hamiltonian} operator $\hat H(t)$ which is a sum of $J$ time-dependent \emph{control} terms and $N(N-1)/2$ two-body \emph{interaction} terms $V_{ij} = V(\mathbf r_{ij})$ that depend on the relative positions $\mathbf r_{ij}$ of each pair of qubits $i, j$ in $\{1, \ldots, N \}$:
\be{eq:hamiltonian}
    \hat{H}(t) = \sum_{j=1}^J \hat H_{\text{ctrl}}^{(j)}(t) + \sum_{i> j} V_{ij} \hat P_i \hat P_{j}.
\ee
with $\hat P_i$ a local operator constructed with Pauli matrices\footnote{The Pauli matrices are
$
\hat \sigma^x = \begin{pmatrix}
    0&1\\
    1&0
  \end{pmatrix}, 
\hat \sigma^y =
  \begin{pmatrix}
    0& -i \\
    i&0
  \end{pmatrix}, \text{ and }
\hat \sigma^z =
  \begin{pmatrix}
    1&0\\
    0&-1
  \end{pmatrix}
$. We represent their action on the $i^{th}$ site of the quantum system by a Kronecker product of matrices: $\hat P_i = \mathbb I\otimes \cdots \otimes \hat P \otimes \cdots \mathbb I$, for $\hat P\in \{\mathbb I, \hat \sigma^x,\hat \sigma^y,\hat \sigma^z \}$ occupying the $i$-th entry . Composite operators such as $\hat P_i \hat P_j$ are then formed via the matrix product.}
and  $\hat H_{\text{ctrl}}^{(j)}(t)$ the $j$-th control term, a time-dependent function representing the action of the experimental parameters affecting the system, usually constructed using waveform generators and applied through laser or microwave fields.

The solution of the differential equation (\ref{eq:schro}) can be obtained by a variety of methods \cite{higham2008functions}: one can approximate the time-evolution operator by $\mathcal{U}(t) \approx \prod \hat U(\delta t)$, with $\hat U(\delta t) = \exp(-i \hat H \delta t)$ for small time intervals $\delta t$, so that $\hat H$ can be considered time independent during $\delta t$. However, computing the exponential of an operator can be prohibitive for large systems \cite{moler2003nineteen}. One can also expand $U(\delta t)$ into a product of simpler terms up to a desired degree of accuracy, depending on the form of $\hat H$~\cite{Barthel2020}. In addition, one can try to numerically solve the full differential equation, such as by using Runge-Kutta methods \cite{hairer2010solving}. Another family of methods consists of implementing an algorithm simulating the time evolution in a quantum computer, for example, by decomposing the exponential into evolution blocks using the Trotter product formulas \cite{cirstoiu2020variational, pastori2022characterization, childs2021theory}, or by using tools from quantum machine learning \cite{caro2022out, gibbs2022dynamical}. One could also try to \emph{learn} the final state vector using tools from machine learning \cite{vicentini2022netket}, or apply Monte Carlo techniques \cite{aoki2014nonequilibrium, de2017dynamics, plenio1998quantum}.

Finally, a very successful approach is given by tensor network techniques, which is the approach we choose to implement as part of our emulation framework. For readers unfamiliar with tensor networks, we briefly review them below. An in-depth detailed description can be found for example in~\cite{Cichocki2016, Cichocki2017, bridgeman2017hand, Schollwck2011}.

\subsection{Representing quantum states and their evolution using tensor networks}\label{sec:tdvp}

Simualting time evolution of non-equilibrium interacting systems is challenging due to the growth of the entanglement entropy with time~\cite{latorre2007entanglement}. Tensor network operations are a matter of linear algebra, and the size of the matrices involved scales with the entanglement of the system under consideration. Different types of tensor networks exhibit different scaling, depending on the geometry and dimension of the physical system~\cite{VerstraetePEPS, Shi2006, Tagliacozzo2009, orus2014tn}, however, for all of these methods, the matrices involved eventually become prohibitively large. Furthermore, only some existing time evolution algorithms can efficiently encode the dynamics of long-range Hamiltonians \cite{Zaletel2015, Paeckel2019}, which are necessary to simulate two-body interactions~\eqref{eq:hamiltonian} with Hamiltonians contains terms of the form $V_{ij} \sim 1/|\mathbf r_{ij}|^\alpha$.

In order to simulate a quantum many-body system \eqref{eq:quantum_state}, we encode the wave function into a \emph{matrix-product state} (MPS)~\cite{Schollwck2011, Oseledets2011}. MPS allow efficient representations of the $N$-order tensor $c_{s_1\cdots s_N}$ as a product of $N$ smaller tensors $A[\sigma]^{s_\sigma}_{i_{\sigma-1}i_\sigma}$:
\be{eq:mps}
|\psi \rangle = \sum_{\{s\}}\sum_{\{i\}} A[1]^{s_1}_{ i_1} A[2]^{s_2}_{i_1i_2}  \cdots A[N]^{s_N}_{i_{N-1}} |s_1  \cdots s_N\rangle,
\ee
where each tensor index $i_\sigma \in {i}$ runs from $1$ to $\chi_\sigma$. The bond dimension $\chi_\sigma$ controls the size of each $A[\sigma]$, which determines the computation complexity of the simulation \cite{latorre2007entanglement}. 

One of the most successful algorithms able to treat long-range interactions, while maintaining a sufficiently small bond dimension, is the Time-Dependent Variational Principle (TDVP) \cite{haegeman2016unifying}. TDVP constrains the time evolution to the MPS manifold $\mathcal{M}(\psi[A])$ with a given $\chi$ by projecting the right hand side of the Schr\"{o}dinger equation onto the tangent space $T\mathcal M (\psi[A])$ :
\be{eq:proj_schro}
\frac{d \ket{\psi[A]}}{dt} = -i \hat \Pi_{T\mathcal M (\psi[A])} \hat H | \psi \rangle.
\ee
The resulting equations \eqref{eq:proj_schro} are nonlinear and couple all degrees of freedom in the MPS. Approaches to solving \eqref{eq:proj_schro} have different ways of controlling accuracy and convergence. In the emulator presented here, we implement 2-site TDVP to deal with the core numerical simulations. Details of the implementation are given in \hyperref[sec:tdvp-details]{Appendix~B}.

\section{Platform architecture}\label{sec:architecture}

In this section we discuss the full architecture of our platform. We describe how the HPC cluster is integrated with cloud services to provide our quantum device emulation. The emulator, which we will refer to as \texttt{EMU-TN}, includes the constraints of a particular QPU as detailed in Section~\ref{sec:hamiltonian-emulator}, implementing the tensor network algorithms of Section \ref{sec:tdvp} as the main numerical backend. Below, we describe the input to the emulator, the pre-processing, as well as post-processing before an output is returned. We finally discuss the cloud infrastructure, including orchestration, scheduling and dispatching.

\subsection{Encoding the dynamics information}

Our platform takes as input an encoded abstract representation of the control parameters $\hat H_\text{ctrl} (t)$. It then performs the required quantum dynamic evolution, applies the measurements to the final state, and returns the readout data to the design tool.

\lil{EMU-TN} includes a \lil{JSON}-parser, which is important to uniformize the information sent to the cluster. We take as initial expression the Hamiltonian in eq.~\eqref{eq:hamiltonian}, with a set of $J$ control fields and the positions $\{\mathbf r_{ij}\}$ of the qubits in the register. Each control field $\hat H_\text{ctrl}^{(j)}(t)$ acts on a subset $I^{(j)}$ of qubits in the register, 
\be{eq:control_field}
\hat H_\text{ctrl}^{(j)}(t) = \Omega^{(j)}(t) \sum_{i\in {I^{(j)}}}  \hat P^{(j)}_i.
\ee
For example, if the $j$-th control is a global field of magnitude $\Omega^{(j)}(t)$ in the $z$-axis, then $\hat P^{(j)} = \hat \sigma_i^z$ and $\hat H_\text{ctrl}^{(j)} = \Omega^{(j)}(t) \sum_{i\in \{ 1 \ldots N \}}  \hat \sigma^z_i$. To parse this information, a design tool will have to specify $\Omega^{(j)}(t)$ as either an array of numbers, the initial and final values plus a total duration, or a generating function. Additionally, locality has to be specified with the indices $I^{(j)}$.

Finally one needs to inform about the initial state preparation and measurement of the state. The initial state is usually experimentally convenient to produce, such as a \emph{product state} indicated typically by a bitstring. Platforms such as neutral atom quantum processors further allow choosing the positions of the qubits, determining the interaction strength between them. More generically, initial state conditions can be provided as labels produced by the design tool. The measurement protocol  can consist of a basis (or a set of bases) where the measurements take place and a number of repetitions of the experiment, in order to generate statistics to estimate observables. This can be specified through a number of \emph{runs} of the job submission, together with a label for the desired basis, as long as it is compatible with the emulated QPU device. 

There is an increasing list of software packages that offer tools for the design and customization of pulse sequences for different types of qubit architectures and quantum processing tasks \cite{teske2022qopt, li2022pulse, alexander2020qiskit, silverio2022pulser, rossignolo2022quocs, goerz2022quantum}. These have been used in the context of the design of quantum algorithms, of quantum optimal control as well as the research of the dynamical properties of many-body quantum systems. A great majority of them use directly or indirectly a state-vector representation of the quantum state, and are available as open source tools for personal use. \texttt{EMU-TN} can be included as an alternative of the solver, thereby extending the types of tasks that can be handled by these design tools.

\subsection{Cloud infrastructure}

We present how the cloud platform dispatches jobs to the classical emulation and quantum computing backends.

\subsubsection{Orchestration and scheduling}
\begin{figure}[ht!]
\centering
\includegraphics[width=0.9\linewidth, clip]{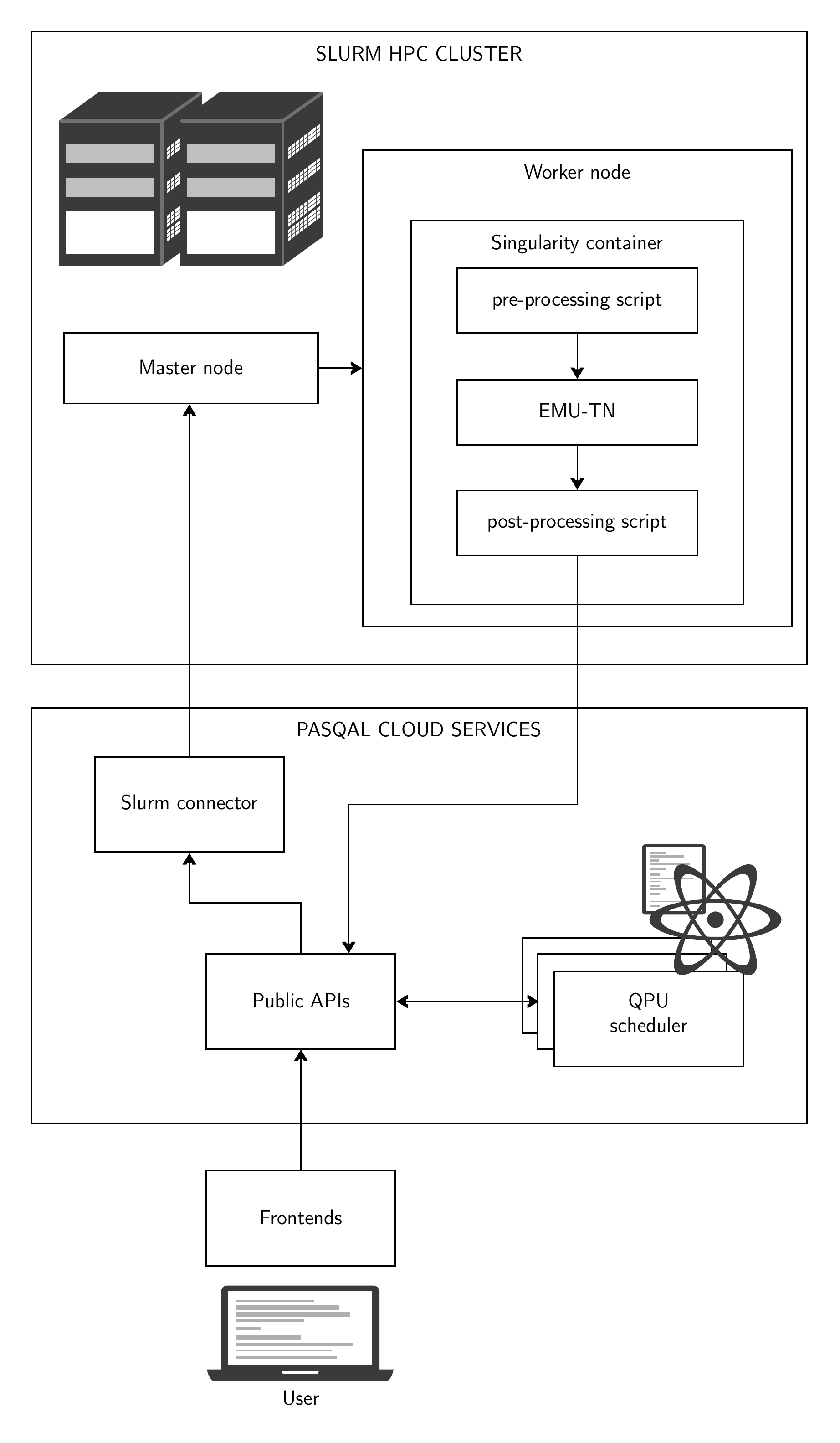}
\caption{Architecture of the cloud platform providing quantum computing as a service.  The cloud services are deployed on a Kubernetes cluster and use a microservice architecture, each microservice exposing a public REST API to end users. The HPC cluster where \texttt{EMU-TN} runs, consists of 10 DGX A100 systems from NVIDIA (See \hyperref[sec:app-cluster]{Appendix A} for more details) and uses Slurm, an HPC cluster management and job scheduling system, for resource management and job submission}
\label{fig:cloud-architecture}
\end{figure}

 The infrastructure, illustrated in Fig. \ref{fig:cloud-architecture}, is split between public cloud services hosted on OVHCloud and orchestrated through Kubernetes and an HPC cluster hosted in a private data center running a Slurm scheduler. We describe below their integration for providing quantum computing as a service. 

Kubernetes~\cite{kubernetes} is a container orchestration system centered around cloud computing, public, private, or hybrid. It simplifies the deployment and operation of containerised applications on a set of resources, as well as the scaling of said resources. Slurm \cite{slurmpaper}, on the other hand, was created with HPC workloads in mind. It provides prioritization aware scheduling, job
queues, topology aware job allocations, reservations and backfill policies. These features led to software like Slurm often being chosen to manage HPC clusters over newer solutions bespoke for other compute paradigms. Slurm submits a general Bash script to run its job, which can also include a sequence of containerised applications.

It is possible to run the same workloads on a Kubernetes pod, either as a long-running service responsible for recieving and executing a job or scheduled using Kubernetes compatible schedulers. However, a downside of these approaches is that it would be challenging to integrate the on-premise HPC cluster with the Kubernetes scheduler. Not scheduling the jobs on the cluster, which would mean just running a job execution service in a Kubernetes pod, would put limitations both on how effectively the resources in the cluster are used, and ultimately on the performance of the HPC applications.

As cloud computing and HPC become increasingly mixed, extensions between Kubernetes and Slurm have recently seen a lot of progress, see for example~\cite{wickberg22slurm, torque_k8s_bridge, slurm_k8s_bridge}. However, these approaches have generally centered around submitting Kuberenetes pods, i.e. containers, to Slurm systems. This involves relying on Kubernetes to schedule jobs. The Kubernetes scheduler does not provide out-of-the-box support to work with often changing and highly heterogeneous environments consisting of several QPU types, CPUs and GPUs. In this work we use a custom scheduler for our pulse sequences. Thus, rather than relying on mechanisms similar to those proposed in \cite{wickberg22slurm, torque_k8s_bridge, slurm_k8s_bridge} a service that directly submits regular Slurm jobs was chosen as this integrates better with our existing hardware and is more similar to the way jobs are scheduled on the QPU.

Moreover, we think that for the foreseeable future hybrid quantum-classical workloads will constitute a large part of quantum jobs. The classical part of these can often be GPU-accelerated, motivating the need to be able to request such compute resources when needed. For example, NVIDIA recently introduced a platform for hybrid quantum-classical computing, known as Quantum Optimized Device Architecture (QODA) \cite{nvidia_qoda}. 

\subsubsection{Job dispatching}

To connect our two clusters, we used the recently introduced Slurm REST API \cite{rini20slurm} to submit jobs to Slurm from the regular Backend service. A ``Slurm connector'' service has been developed to act as a bridge between the existing PASQAL cloud Services and the HPC cluster. The service creates a control script on-demand for the job which is submitted to Slurm as the batch script. This script then takes care of executing all the higher-level logic of the job. The service is also responsible for setting up the job with appropriate resources and inserting all necessary information for future communication with the cloud service.

Each pulse sequence is encoded in a \texttt{JSON} format and sent to PASQAL cloud Services by making an HTTP POST request to the job submission endpoint. The pulse sequence format is the same for QPUs and emulators. We extend the body of the HTTP request with a \texttt{Configuration} argument that allows the user to control the parameters of the emulator, see \hyperref[lst:cloud-sdk]{Code Sample 1}. This design allows executing the same quantum program seamlessly on both QPUs and emulators, while also allowing control of the numerical simulation parameters such as the maximum bond dimension and the number of cores.

The internal logic of the job submission endpoint validates the pulse sequence and the emulation configuration. The validation step checks that the request and the underlying data such as the sequence and emulator configuration are correctly formatted, that the requested resources are accessible to the user and, for a real device, that the sequence is physically valid. It then dispatches the request to the appropriate scheduler. In \hyperref[lst:cloud-sdk]{Code Sample 1} below, we show an example of sending a serialized pulse sequence to PASQAL Cloud Services for execution. To build the sequences we have used \texttt{Pulser} \cite{silverio2022pulser}, an open source package for designing neutral atom QPU sequences. To communicate with cloud services we have used the PASQAL Cloud Services Python SDK~\cite{PCS2021}, which provides helper functions to the end user to request the APIs.

\begin{lstlisting}[caption=Running jobs through the cloud service., label={lst:cloud-sdk}, language=python]
from pulser import Pulse, Sequence, Register 
from pulser.devices import MockDevice
from sdk import Configuration, Endpoints, SDK
import json

# 1. Define Qubit Register
N = 6 # Qubits per side 
reg = Register.square(side=N, spacing=7, prefix="q")

# 2. Define Pulse Sequence and encode to JSON
seq = Sequence(reg, MockDevice)
seq.declare_channel(
    name="ch0",
    channel_id="rydberg_global"
)
seq.add(
    Pulse.ConstantPulse(
        duration=1000,  # ns
        amplitude=2, # rad/us
        detuning=-6, # rad/us
        phase=0),
    channel="ch0"
)
encoded_seq = seq.to_abstract_repr() # JSON file

# 3. Send encoded sequence to PASQAL Cloud Services
#   Get API keys at
#   https://portal.pasqal.cloud/api-keys

endpoints = Endpoints(
    core="https://apis.pasqal.cloud/core",
    account="https://apis.pasqal.cloud/account"
)

sdk = SDK(
    client_id="my_client_id", 
    client_secret="my_client_secret", 
    endpoints=endpoints,
)

# Configure Job
my_job = {"runs": 1000} # Number of shots
config = Configuration( # Configuration of emulator
    dt=10.0, 
    precision="normal",
    extra_config={"max-bond-dim": 100}
)

# Create batch on the Cloud service, wait until the execution completes, and get results
batch = sdk.create_batch(
    encoded_seq, 
    device_type="EMU_TN",
    jobs=[my_job],
    configuration=config,
    fetch_results=True,
)
res = [job.result for job in batch.jobs.values()]
\end{lstlisting}

In this context, \lil{EMU-TN} is built as a Singularity container and executed as a Slurm job. The Slurm job consists of the numerical simulation and is preceded and followed by custom-made scripts to communicate with the cloud platform. Since the pulse sequence has been previously validated, the pre-processing script only translates the user configuration into the format required by \lil{EMU-TN}. After the completion of the simulation, the post-processing script collects the results and communicates with the cloud service to ensure the results are correctly stored and accessible by the frontend. This architecture allows the emulators to be stand-alone and also to be easily executable outside the overall platform. It also allows for easy extension for more frequent two-way communication between the platform and the emulator, should it be desirable in the future.


\subsubsection{Other considerations}

Storing bistring results or heavily compressed tensor-states in a columnar database works very well as they can be serialised to short text strings. For larger datasets this can cause performance degradation in the database and eventually the dataset will be too large for the specific database implementation. Thus, object storage may be useful in such cases, such as for the tensor network representations produced by the emulator. Since obtaining such a state is computationally costly, the platform makes this state available as a serialised object that can be de-serialised for further processing as required by the numerical algorithm. 

A second consideration is that not every user request constitutes a feasible computational objective. Some pulse sequences (e.g. those that introduce steep variations) usually require a larger matrix size to simulate. It could also be that the algorithm does not converge to the desired accuracy. There is an unavoidable iterative process where a candidate pulse sequence and quantum system is studied by the user on physical grounds and successive verification with exact diagonalization methods for smaller systems is crucial. Once a promising job is selected, the data is deserialized in the cloud core service where a validation script checks that it describes a valid quantum program, for a given emulated device based on existing QPUs. This ensures that the  pulse sequence is physically implementable on the hardware. The Slurm-connector also does some simple validation of Slurm configuration. 

\section{Applications}\label{sec:applications}

In this section we discuss some applications of the \lil{EMU-TN} framework, motivated by typical tasks that can be performed with a Neutral Atoms Quantum Processor \cite{henriet2020quantum, saffman2010quantum, morgado2021quantum}. The native Hamiltonian will be taken as the following particular form of (\ref{eq:hamiltonian}):
\be{eq:ham_atoms}
    \hat{H}(t) =  \frac{\Omega(t)}{2} \sum_{i=1}^N \hat \sigma^x_i - \delta(t) \sum_{i=1}^N \hat n_i + \sum_{i>j} \frac{C}{|\mathbf r_{ij}|^6} \hat n_i \hat n_{j}
\ee
where $\hat n = \proj{1}{1} = (\mathbb I - \hat \sigma^z)/2$ is the projector into the excited state and $C$ is a constant that depends on the properties of the targeted excited state. After the quantum evolution and readout, one obtains a bitstring of 0's and 1's representing excited and ground-state qubits, respectively. To calculate the expectation value of a given observable, many cycles need to be performed in order to generate enough statistics. The output of the emulator provides bistrings, which simulates the actual output of the QPU.

We remark that \lil{EMU-TN} can be applied to other Hamiltonians, for example long-range ``XY''  Hamiltonians that can be constructed with neutral atom~\cite{de2017optical} and trapped ion~\cite{jurcevic2014quasiparticle} arrays.

\subsection{1D Regular Lattice: Rydberg Crystals}

One notable property of the Hamiltonian \eqref{eq:ham_atoms} is that, in certain regimes, the simultaneous excitation of two neighbouring atoms is suppressed. This phenomenon is called Rydberg blockade \cite{Jaksch2000, browaeys2020many}, and it takes place when the interaction term $C/r^6$ exceeds the Rabi frequency $\Omega$ (it is therefore distance dependent). The Rydberg blockade mechanism is the main source of entanglement in neutral atom systems, since the maximally entangled state $\ket{01}+\ket{10}$ becomes energetically favored over the product state $\ket{11}$.

\begin{figure}[ht!]
    \centering
    \includegraphics[width=0.6\linewidth]{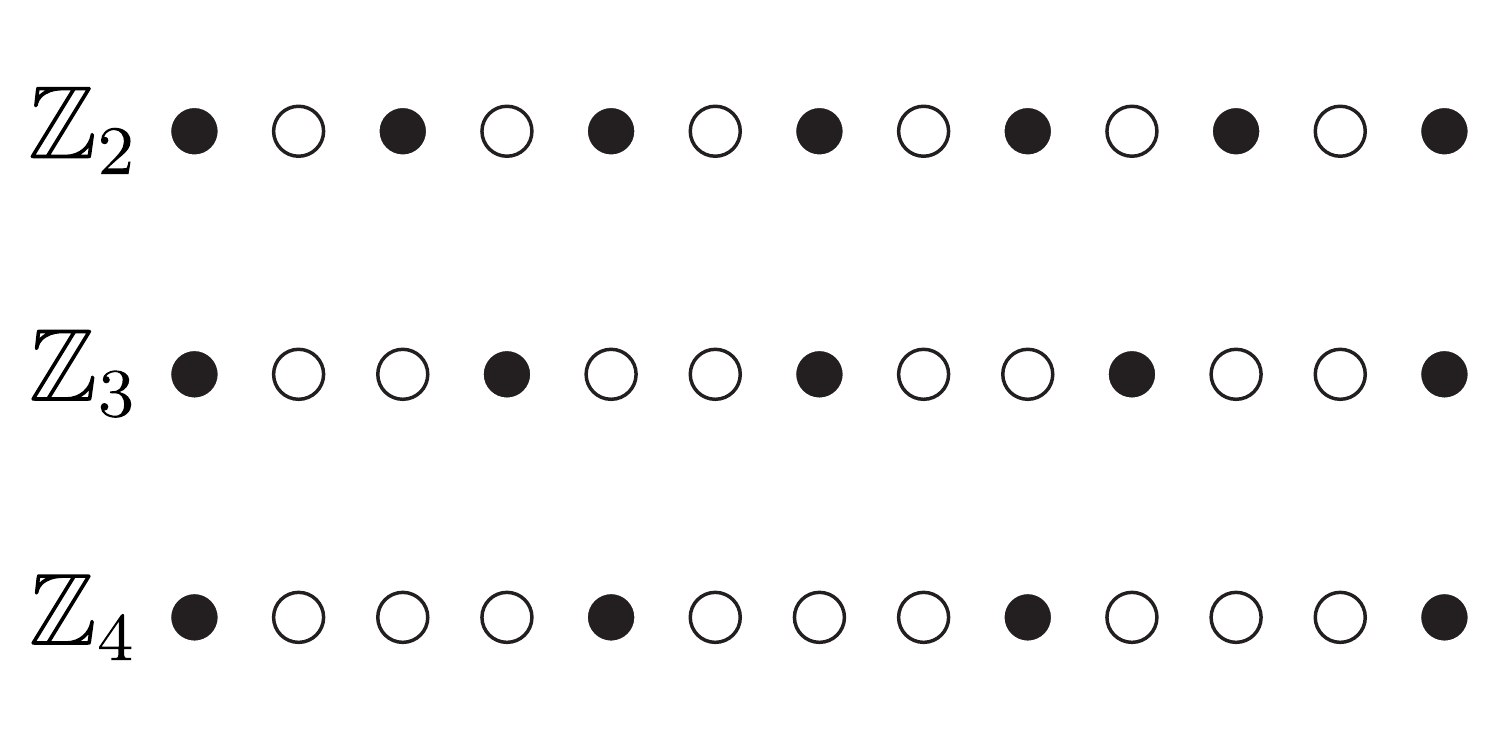}
    \caption{Examples of 1D Rydberg crystals for a chain of 13 atoms. A solid circle represents an excited atom, while an empty one represents an atom in the ground state. The $\mathbb{Z}_2$, $\mathbb{Z}_3$ and $\mathbb{Z}_4$ orders are  characterized by excitations separated by respectively 1, 2, or 3 atoms in the ground state.}
    \label{fig:rydcry_z234}
\end{figure}

The Rydberg blockade has been used in \cite{Bernien2017} to create highly regular excitation patterns in 1D chains of neutral atoms. These structures are known as Rydberg crystals and, for experimentally reasonable values of atomic spacing and Rabi frequency, they can be prepared in such a way as to display $\mathbb{Z}_n$ order for $n=2,3,4$. The situation is depicted schematically in Fig. \ref{fig:rydcry_z234}.

\begin{figure}[ht!]
    \centering
    \includegraphics[width=0.8\linewidth,trim= 7cm 7cm 9cm 8.5cm, clip]{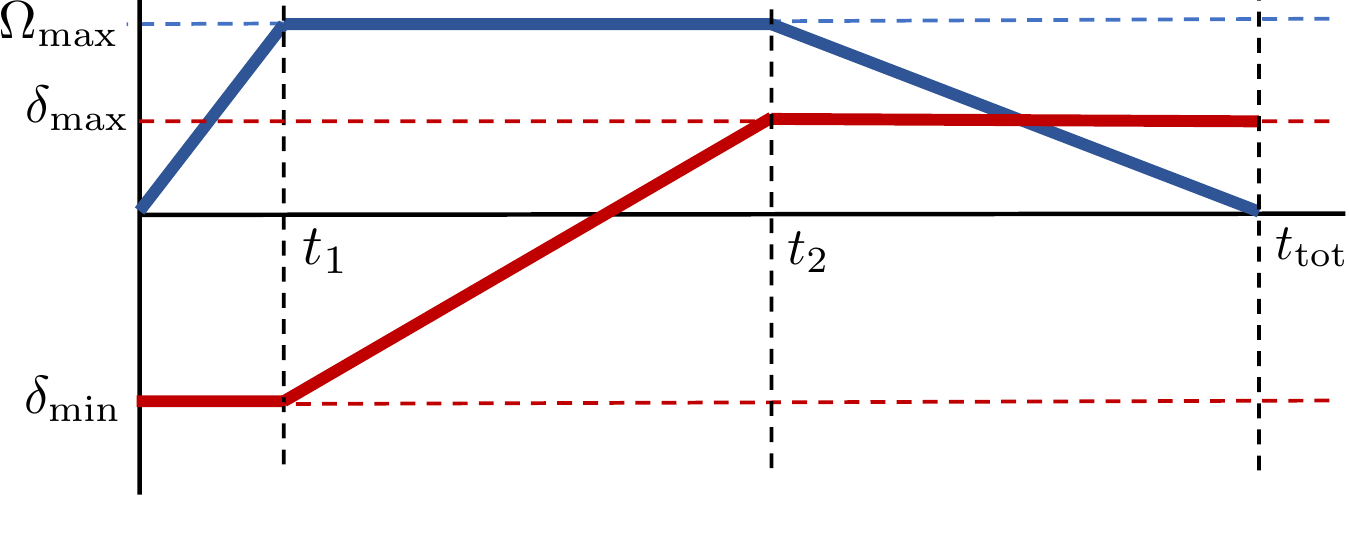}
    \includegraphics[width=0.9\linewidth]{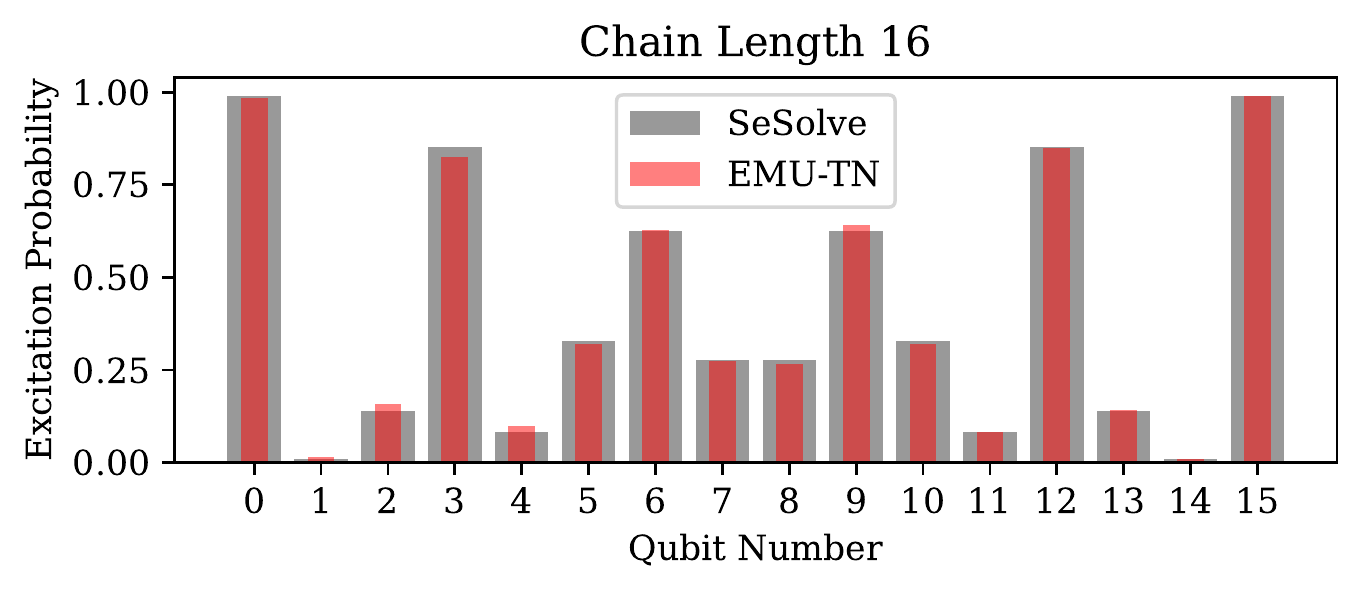}
    \includegraphics[width=0.9\linewidth]{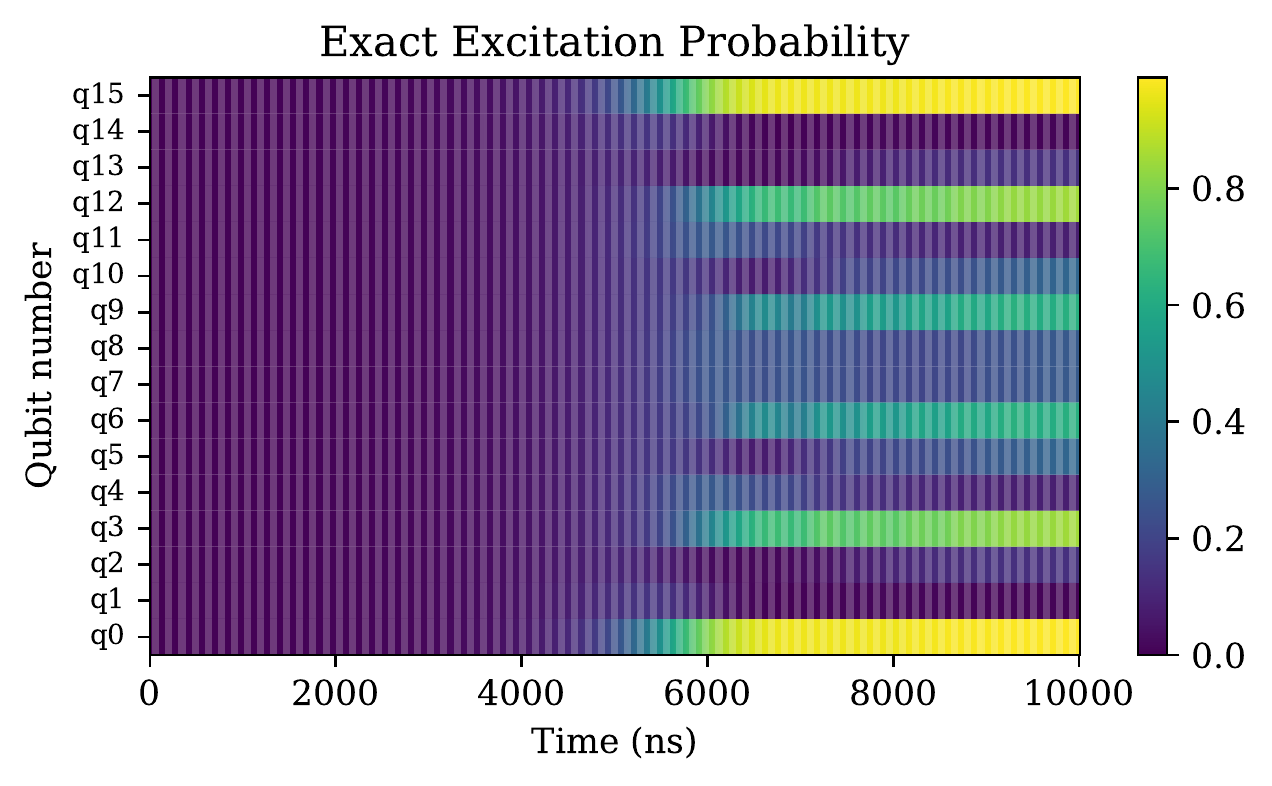}
    \includegraphics[width=0.9\linewidth]{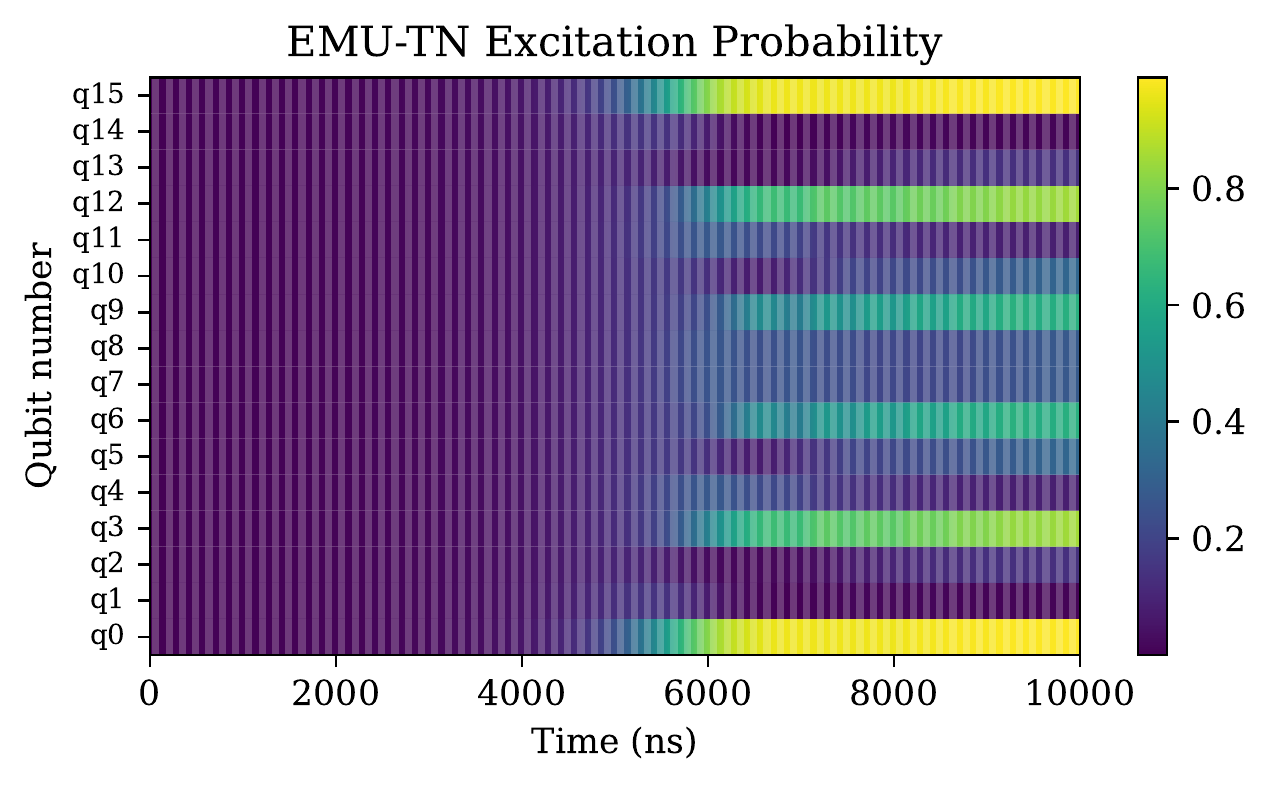}
    \caption{Comparison between an exact Schr\"odinger equation solver (using \texttt{QuTiP} \cite{johansson2012qutip}) and the emulator (\texttt{EMU-TN}) for $\mathbb{Z}_3$ Rydberg crystal preparation on a 1D chain of 16 atoms placed 4 $\mu m$ apart. The driving sequence represents a pulse of $t_{tot} = 10 \mu s$ with $\Omega_\text{max}=6.3 \ \text{rad}/\mu s$. The histogram represents the excitation probability for each qubit in the final state, while the heatmaps represent the evolution over time of the excitation probability obtained with the exact solver and \texttt{EMU-TN} respectively.}
    \label{fig:rydcry16}
\end{figure}

We first present the results of a test simulation on a chain of 16 atoms. For the pulse sequence shown in the top part of Fig. \ref{fig:rydcry16} with $\Omega_\text{max}=6.3 \ \text{rad}/\mu s$ and a spacing of 4 $\mu m$, we expect the final ground state of the system to be in the $\mathbb{Z}_3$ order. The excitation probability for each qubit in its final state is also reported for both \lil{EMU-TN} and an exact Schr\"odinger equation solver in Fig. \ref{fig:rydcry16}, together with a heatmap representing the time evolution of the excitation probability for each qubit given by the exact solver and \lil{EMU-TN} respectively.

\begin{figure}[ht!]
    \centering
    \includegraphics[width=0.9\linewidth]{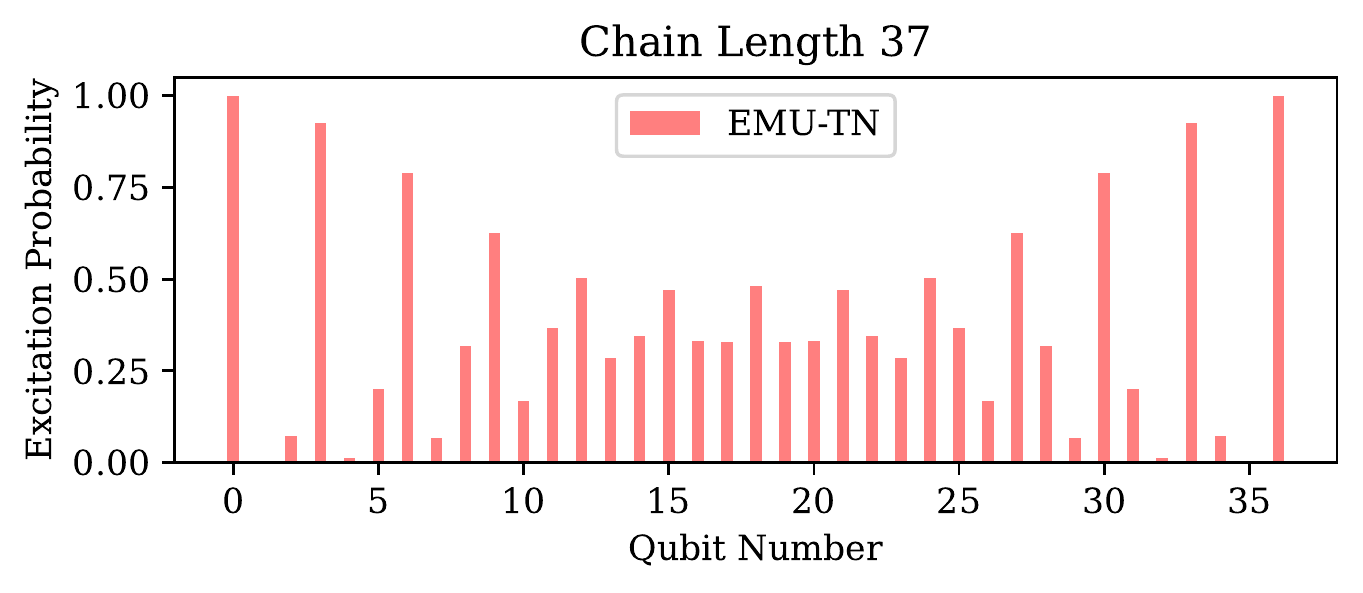}
    \includegraphics[width=0.9\linewidth]{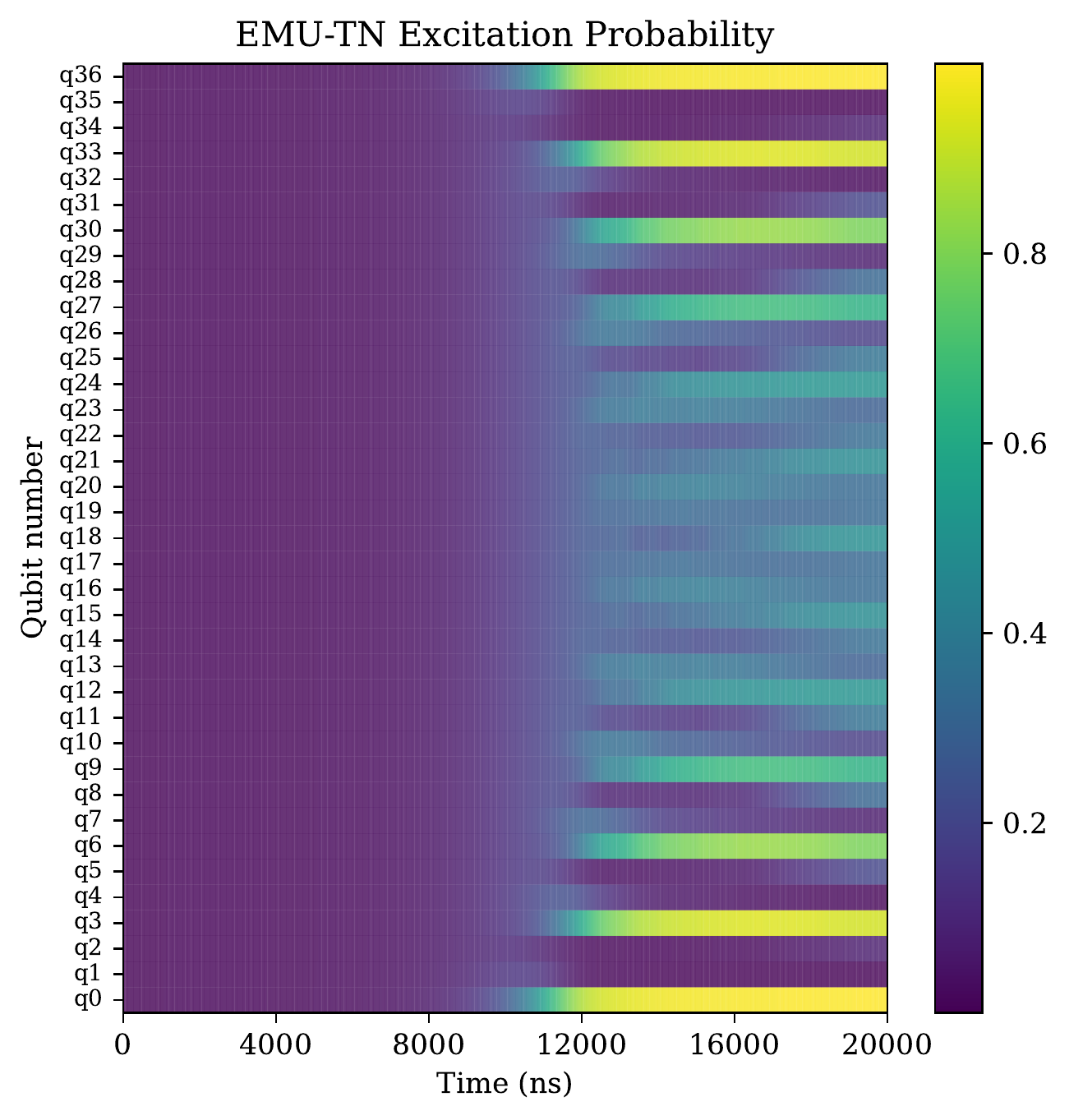}
    \caption{Same simulation as Fig. \ref{fig:rydcry16}, but with a chain of 37 qubits and a pulse of 20 $\mu s$.}
    \label{fig:rydcry37}
\end{figure}

Next, we present the results for the same simulation but with a chain of 37 atoms, which is intractable by an exact solver. The pulse in this case is chosen to be twice as long in order to allow correlations to spread from the borders all the way to the center of the chain and observe a clearer alternation of the excitations. The excitation probability in the final state and its time evolution are reported in Fig. \ref{fig:rydcry37}.

\subsection{2D Regular Lattice: Antiferromagnetic State Preparation}

One of the most common tasks one can try to simulate is the preparation of particular states with a regular register. This has been used in milestone implementations in programmable neutral-atom arrays of hundreds of qubits \cite{scholl2021quantum, ebadi2021quantum}.

\begin{figure}[ht!]
    \centering
    \includegraphics[width=0.49\linewidth, clip]{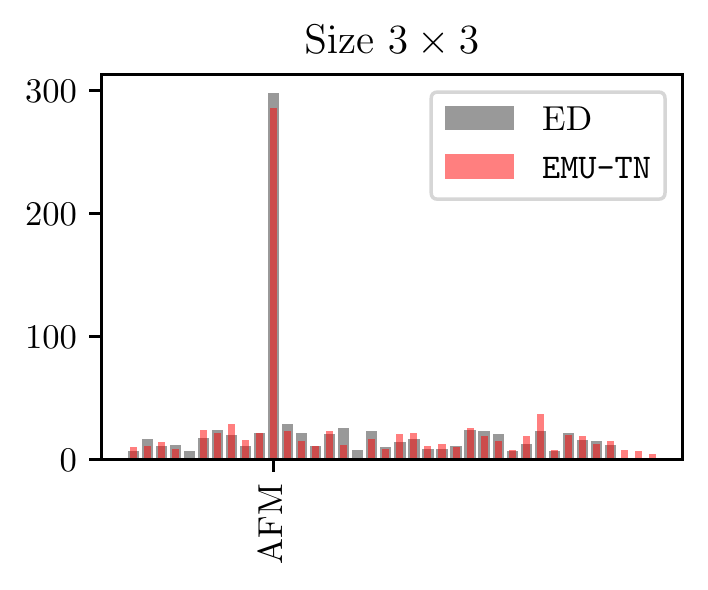}
    \includegraphics[width=0.49\linewidth, clip]{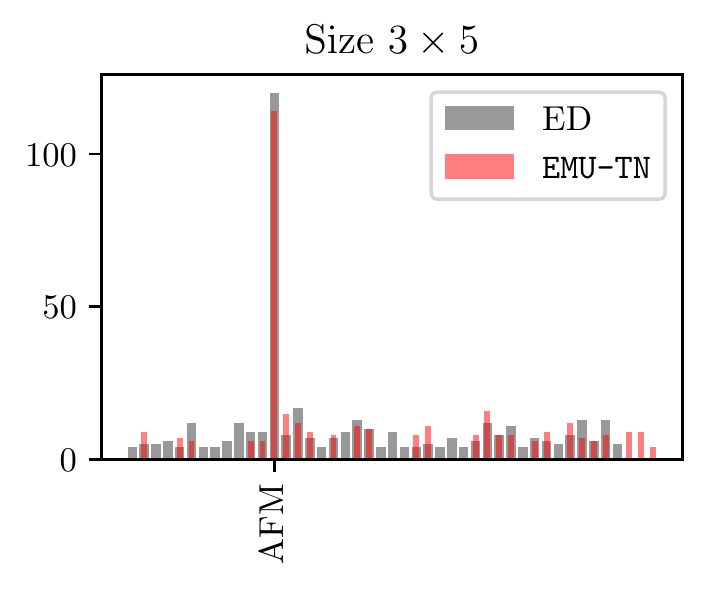}
    \includegraphics[width=0.49\linewidth, clip]{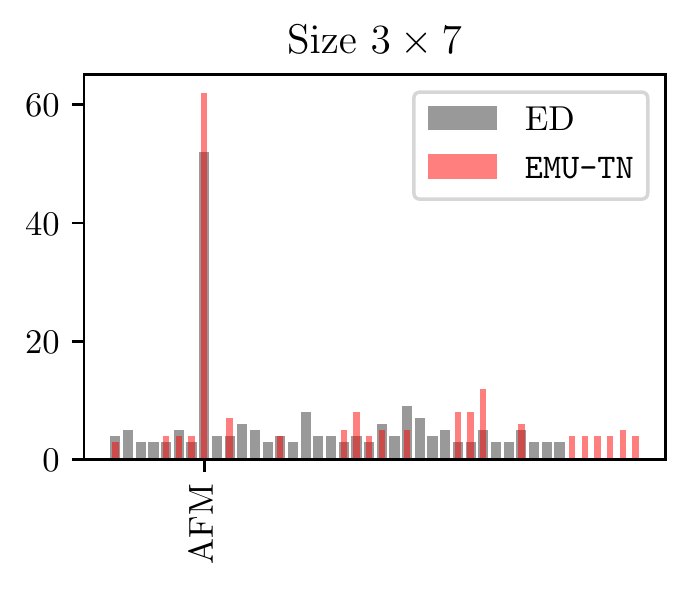}
    \includegraphics[width=0.49\linewidth, clip]{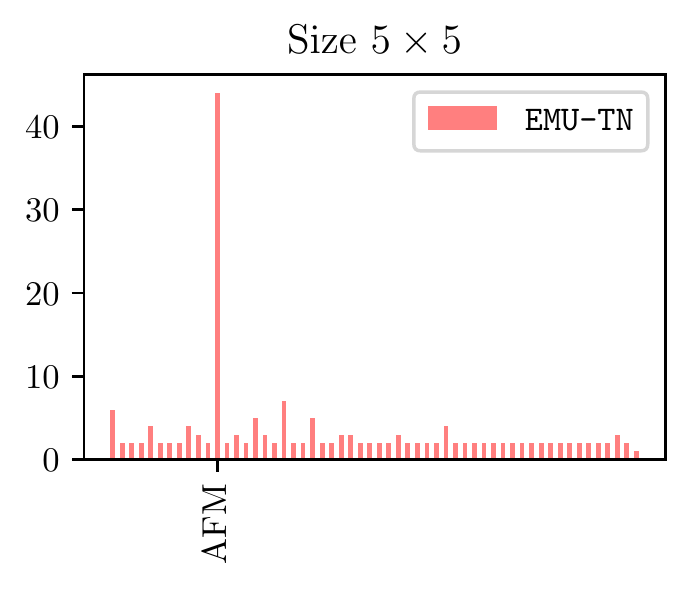}
    \includegraphics[width=0.49\linewidth, clip]{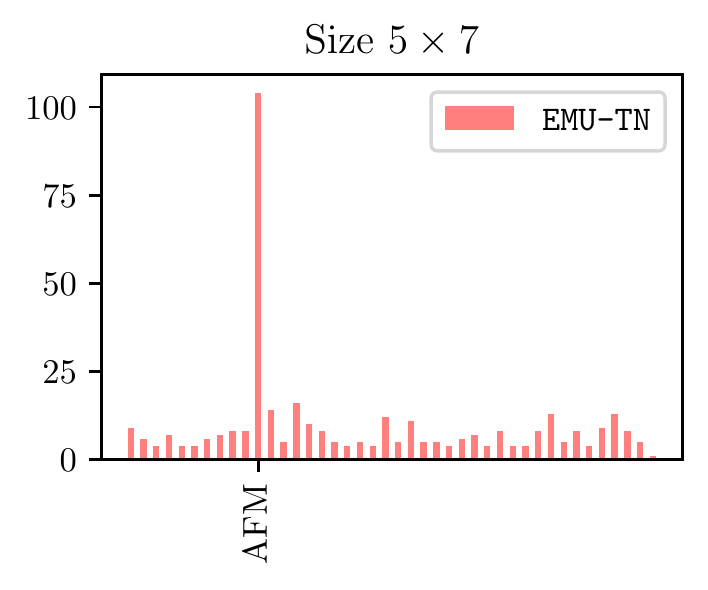}
    \includegraphics[width=0.49\linewidth, clip]{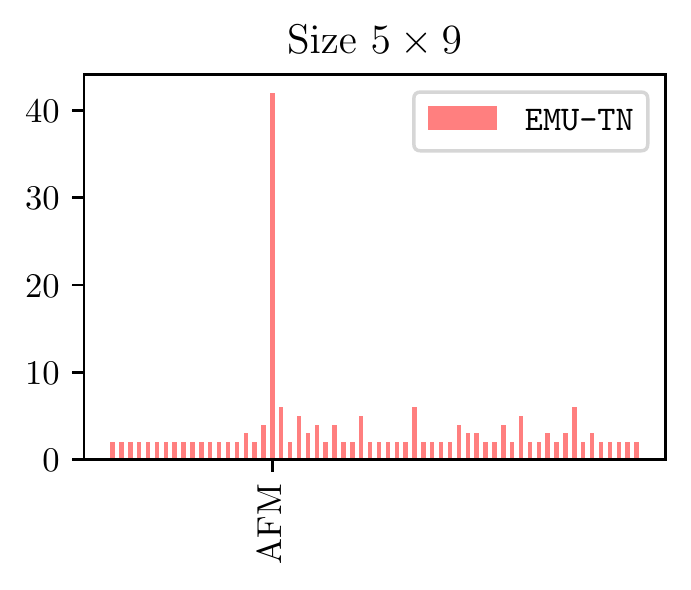}
    \caption{Comparison of sampling results ($10^3$ samples) from the prepared quantum state between exact diagonalization (ED) and the emulator (\texttt{EMU-TN}) for increasing array sizes. The driving sequence (same structure as the one in Fig. \ref{fig:rydcry16}, above) represents a pulse of $t_\text{tot}=2.7 \mu s$ moving in the phase diagram of the Hamiltonian towards the antiferromagnetic phase. The $5\times 7$ and $5\times 9$ arrays required $10^4$ samples to resolve better the peaks (1\% and 0.4\% respectively), since the short pulse breaks adiabaticity for large arrays.}
    \label{fig:AFM_hist}
\end{figure}

\begin{figure}[ht!]\label{fig:benchmarks}
    \centering
    \includegraphics[width=0.9\linewidth]{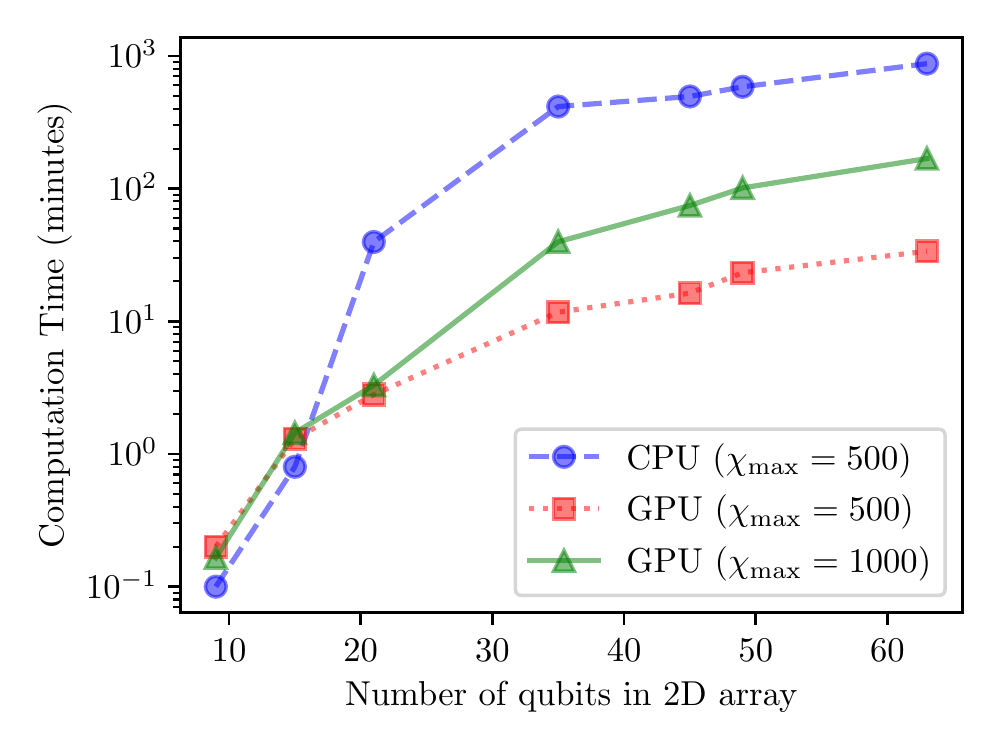}
    \caption{Computation time for increasing array sizes. We compare only CPU use and then an added GPU backend, for different maximal bond dimension, $\chi_{\text{max}}$.}
\end{figure}

A typical pulse sequence (Fig. \ref{fig:AFM_hist}, above) represents a path through the phase diagram in the thermodynamic limit that ends in a point of the expected antiferromagnetic phase, which has been analytically studied before. We present the results from the sampling of the evolved MPS as well as from a straightforward implementation of exact diagonalization solved numerically on a local computer. Using exact diagonalization it is possible to run simulations just above the 20-qubit range, but it soon becomes impractical once the number of qubits increases. On the other hand, adjusting the bond dimension of the tensor network algorithm, one can aim to explore the behavior of sequences in the range of 20 to 60 qubits in a comparably short time. Moreover, the flexibility of the cluster approach allows for parallelization of tasks which provide information about parameter sweeps without the user having to spend time adapting their code. In Figure \ref{fig:benchmarks}, we include information about the elapsed time of the performed simulations, for different array sizes and bond dimensions, with and without access to GPUs.

\subsection{Maximum Independent Set on Unit Disk graphs}

We present a typical use case of neutral atom devices for two-dimensional registers. The possibility of placing atoms in arbitrary 2D configurations allows to solve certain classes of hard graph problems on the QPU. An interesting application is solving the Maximum Independent Set (MIS) problem on Unit Disk (UD) graphs \cite{pichler2018}.

A graph is an object with nodes and connections between nodes. A UD graph is a graph where two nodes are connected if an only if they are closer than a certain minimal distance. By representing the nodes of a graph with neutral atoms and matching the Rydberg blockade radius with the UD graph minimal distance, one can establish a direct mapping where a connection exists between atoms that are within a blockade distance of each other. The quantum evolution of such a system is naturally restricted to those sectors of the Hilbert space where excitations of connected atoms are forbidden. These configurations translate to independent sets of a graph, {\sl i.e.} subsets of nodes that are not directly connected to each other. Driving the quantum evolution in such a way as to produce as many excitations as possible, one would obtain with high probability as the outcome of a measurement an independent set of high cardinality, representing a good candidate solution to the MIS problem.

\begin{figure}[ht!]
    \centering
    \includegraphics[width=0.9\linewidth]{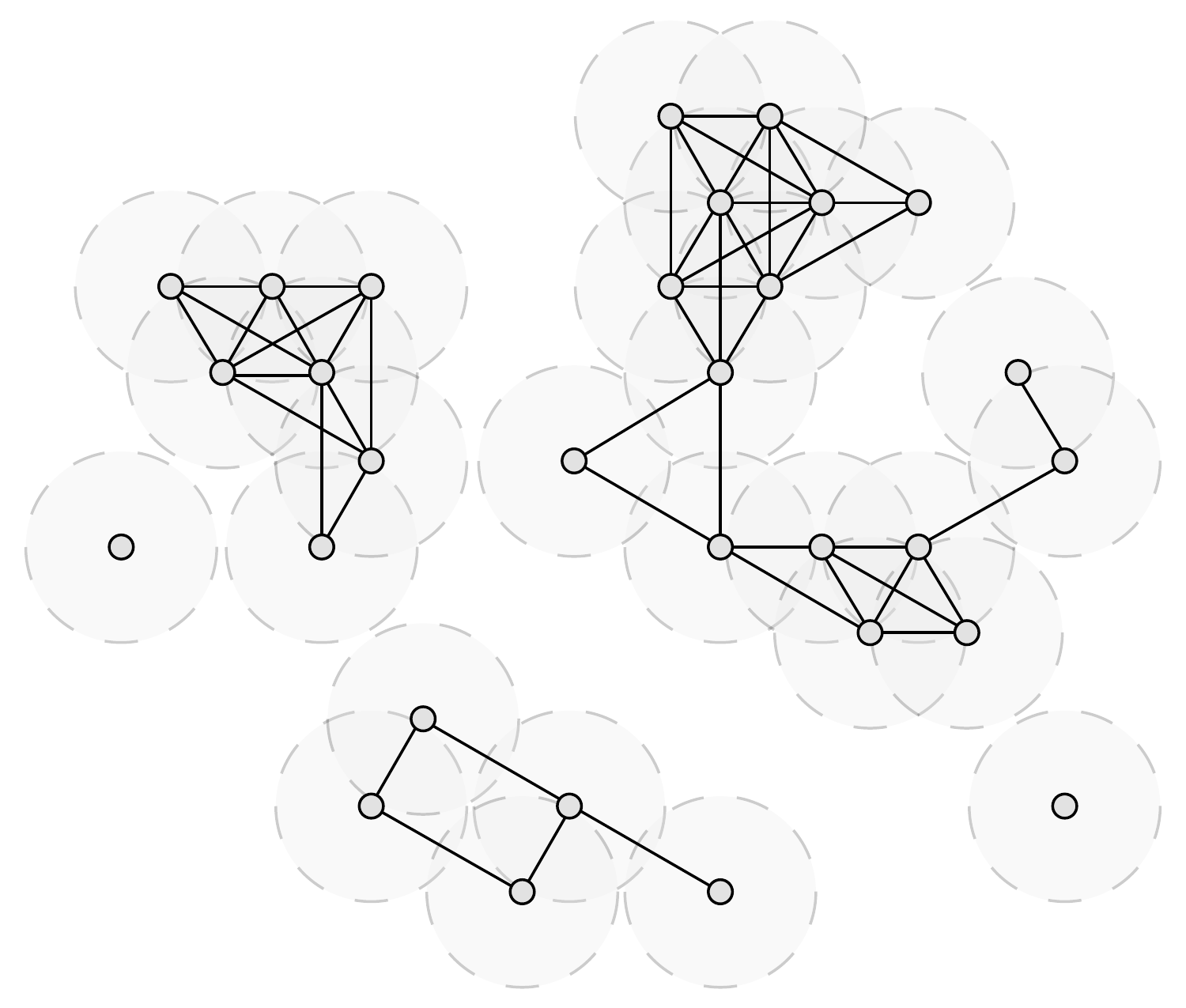}
    \includegraphics[width=0.9\linewidth]{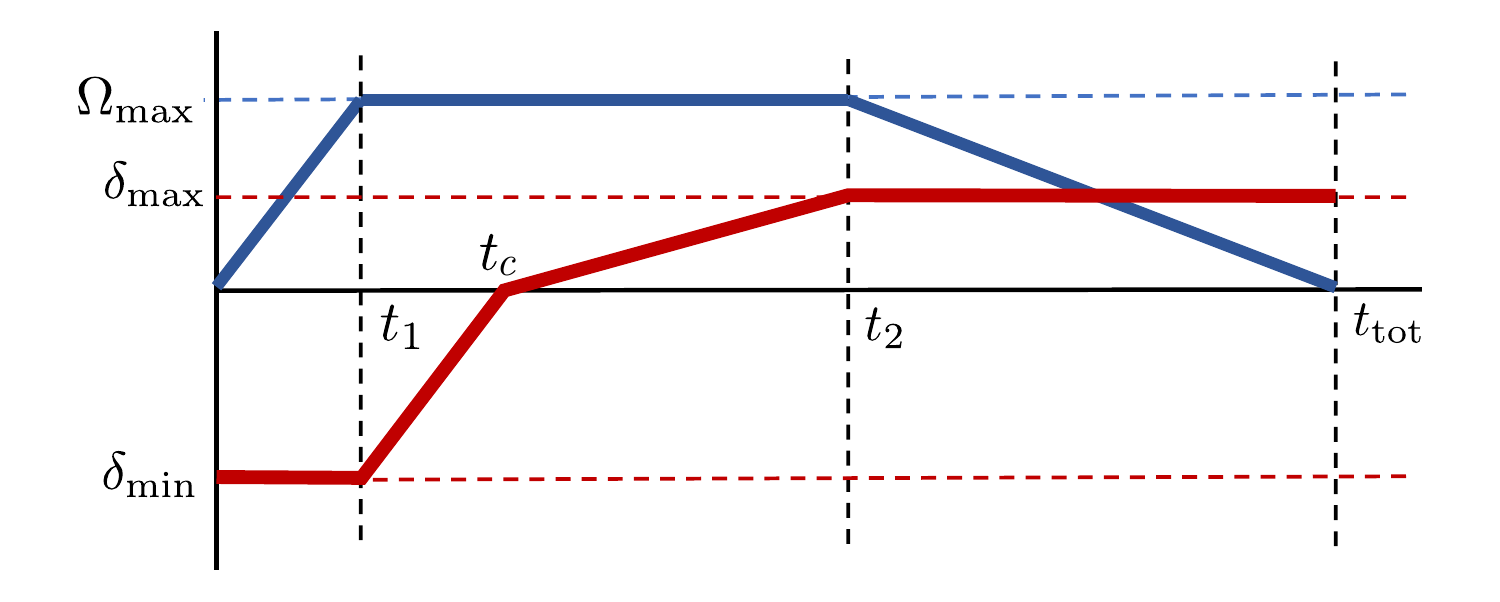}
    \includegraphics[width=0.9\linewidth]{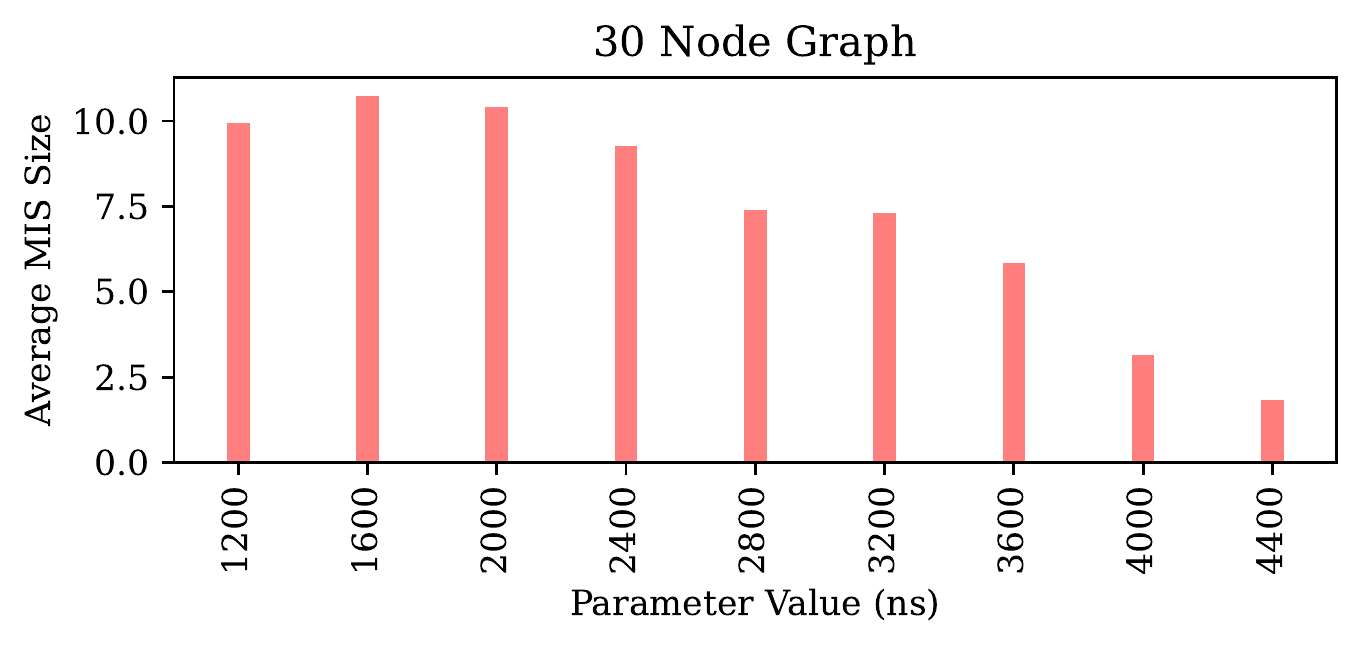}
    \caption{Parameter sweep. We arrange 30 qubits in 2D (above), where the dashed circles show half the blockade radius. The pulse sequence (middle, $t_\text{tot} = 5 \mu s$) contains a tunable parameter $t_c$ controlling the time when the detuning goes from negative to positive. We sweep in parallel for 9 values of $t_c$. The results (below) show the average size of the MIS found for each value of $t_c$. The entire sweep, took about 2 hours of cloud compute time.}
    \label{fig:mis30}
\end{figure}

The best pulse sequence to find the MIS of a graph is based on the same adiabatic pulses already shown in Figs. \ref{fig:rydcry16} and \ref{fig:AFM_hist}, partly since Rydberg crystals and antiferromagnetic states can be seen as the MIS of regular 1D and 2D lattices. As shown in \cite{ebadi2022}, however, for arbitrary graphs it is often necessary to optimize the pulse shape in order to find the MIS with a high enough probability. We present in Fig. \ref{fig:mis30} a one-parameter family of pulses where the parameter $t_c$ controls at which time the detuning goes from negative to positive. Rather than sweeping over the free parameter sequentially, one can send multiple simultaneous jobs for each value to be tested, and finally select the parameter giving the best result. Parallelizing this task is a direct advantage of the cluster infrastructure, and the cloud service provides a dedicated interface in which all jobs can be organized and retrieved, saving large amounts of time. We performed a sweep over 9 parameters for a pulse of 5 $\mu s$ on a graph with 30 nodes. The results are shown in Fig. \ref{fig:mis30}, below, where the optimal zero-crossing for the detuning is found to be at around 1.6 $\mu s$.

While the example shown here is centered on pulse parameters, other tasks suited for parallelization are also implementable, e.g. averages over noise and disorder realizations. Being able to quickly set up and run studies of this kind for systems of many qubits opens the door for creative algorithm design that is at the same time compatible with a QPU. We remark that in 1D systems, the TDVP algorithm has been used to study dynamics for $\mathcal{O}(10^2)$ spins \cite{yang2020time, li2022time}.

\section{Discussion}

Due to the rising interest in developing and using quantum processing units integrated in a modern computational infrastructure, we envision the emergence of workflows requiring intensive use of emulators. They are not just a placeholder for the actual quantum device, but serve for diagnostics, resource estimation and management. Given the current price and availability of QPUs, emulation is crucial in the initial phases of an application, providing a faster feedback loop during the R\&D phase, as well as enforcing a realistic level of constraints based on a given quantum computing architecture. This is also relevant for designing upgrades to a quantum device, as the relevant figures of merit for expected behavior can be compared quickly.

By providing a seamless interface between emulators running locally, emulators of large systems in a HPC environment and the QPUs, the researcher can distribute the most convenient compute option. By further centralising this in a workflow and allowing the scheduler to manage the optimal software resource one can both reduce the total time and cost to solution. Workflows will also have to take into account whether the computational task should be executed in the shortest walltime, the most energy efficient way, the most cost-effective way or some weighted combination. The computational community has taken great strides in workflow management, and the ever increasing use of machine learning has created new fields like MLOps (Machine Learning Operations). It is expected that the quantum computing community strives for such state of the art workflows, and to be able to integrate into existing ones, to achieve real industrial advantage. This will include not only execution of algorithms, but also providing applicable training, and testing.

In this paper we described the components at the basis of such workflows. By leveraging tensor network algorithms for quantum dynamics that can exploit the resources provided by an HPC cluster, we discussed how a cloud on-demand service can make use of a cluster backend to simplify access to the required computational operations. The numerics can be easily adapted to different architectures and devices, separating the role of design tools from the computation of quantum dynamics. Other techniques for the study of dynamical quantum systems can be adapted as well: future objectives include considering the low qubit ($< 20$) range, where state-vector evolution methods allow implementing a microscopic model of hardware noise effects and other state-preparation and readout errors \cite{de2018analysis}. This includes implementing parallelization routines on the algorithm management side, controlled exploitation of the cluster resources for heavy computational loads and efficient solvers for other dynamical equations (quantum master equations and stochastic Schr\"odinger equation). A suite of benchmarks for common tasks in the low qubit range is also desirable for understanding speed bottlenecks, and will be the subject of a follow up work. In the case of larger arrays, the use of tensor network structures like PEPS or Tree Tensor Networks, provide other convenient representations for two-dimensional systems and their application to a time evolution algorithms is under active study \cite{zaletel2015time, bauernfeind2020time, rams2020breaking, kloss2020studying, vanhecke2021symmetric}

Developing workflows is much easier with a fully fledged cloud backend to orchestrate the tasks. This enables creating workflows that automatically assess the validity of the submitted pulse sequence for the QPU. Simulating the dynamics from a particular pulse sequence requires attention to tuning and understanding how to set the algorithm parameters. The cloud platform can provide the user suggestions for future hyperparameters based on live monitoring of the simulation. Finally, the worker nodes on the cluster have large amounts of local storage in a so-called ``scratch'' directory. This allows storing large amounts of data cheaply on a temporary basis. A future extension of the platform could allow subsequent job-steps to use previous results, recycling previous computational work for new tasks.

We believe the platform presented here can greatly improve reproducibility and benchmarking of research in the developing quantum industry, as well as allowing a broader scope of researchers, engineers and developers to understand, explore and design the behavior of quantum systems and quantum devices. 

\begin{acknowledgements}
We thank Lo\"ic Henriet, Caroline de Groot and Matthieu Moreau for valuable conversations, suggestions for this manuscript and collaboration in related work.
\end{acknowledgements}

\newpage


\bibliographystyle{unsrtnat}
\bibliography{main.bib} 

\section*{Appendix}

\subsection{Computation Cluster}\label{sec:app-cluster}

As an example of a modern HPC cluster infrastructure in which the algorithms mentioned in the main text are expected to be executed, we describe PASQAL's computation cluster. Note that the amount of computing power of different providers varies widely, as does their availability and cost. This is a crucial element to consider when intensive computational tasks are needed.

Both CPUs and GPUs can be used to compute complex scientific algorithms, although they are often used for different types of tasks. CPUs are generally more versatile and can handle a wide range of tasks, including running the operating system and other software, managing memory and input/output, and performing complex calculations. They are well suited for tasks that require sequential processing, such as executing instructions in a specific order or following a set of rules. GPUs, on the other hand, are specialized processors that are designed to handle the large number of calculations needed to render images and video. They are particularly well suited for tasks that can be parallelized, meaning that they can be broken down into smaller pieces consisting of identical operations on potentially different data. 
\newline \\
\emph{GPU.-} The NVIDIA A100 GPU delivers subsequent acceleration for a wide range of applications in AI, data analytics, and high-performance computing (HPC). A particularly parallel algorithm can efficiently scale to thousands of GPUs or, with NVIDIA Multi-Instance GPU (MIG) technology, be partitioned into seven GPU instances to accelerate workloads of all sizes. Furthermore, third-generation Tensor Cores accelerate every precision for diverse workloads. At PASQAL, we currently have 80 A100 NVIDIA GPUs distributed into 10 DGX A100 NVIDIA systems.
These servers are linked to each other through Mellanox's proprietatry 200Gbps Ethernet fibre optics connections. Each DGX A100 system is itself comprised of 2 Dual 64-core AMD EPYC 7742 CPUs and 8 NVIDIA A100 SXM4 GPUs. Each GPU is connected through NVLINK to each other, with offers a GPU memory pool of 320 GB of shared GPU memory for one node, in addition to the 1 TB of RAM memory for the AMD CPUs.
\newline \\
\emph{CPU.-} With two 64-core EPYC CPUs and 1TB of system memory, the DGX A100 boasts high performance even before the GPUs are considered. The architecture of the AMD EPYC “Rome” CPUs is outside the scope of this article, but offers an elegant design of its own. Each CPU provides 64 processor cores (supporting up to 128 threads), 256MB L3 cache, and eight channels of DDR4-3200 memory (which provides the highest memory throughput of any mainstream x86 CPU). The 15TB high-speed NVMe SSD Scratch Space allows for very fast data read/write speeds which definitely impacts total processing speeds when large data needs to be uploaded from disk to RAM memory at different steps of the computation.
In addition to the on-node scratch space the cluster has a 15TB distributed file system providing the users home directories as well as software and containers. Using Slurm's partitions we separate internal, external and production usage coming from the cloud service herein.

\subsection{TDVP implementation details}\label{sec:tdvp-details}

For the applications shown in the Section~\ref{sec:applications}, Tensor network calculations are implemented by using the \texttt{ITensor.jl}~\cite{itensor} and \texttt{ITensorTDVP.jl} \cite{fishman2022tdvp} packages written in \texttt{Julia} language \cite{bezanson2017julia}.
As a time evolution algorithm  we chose 2-site TDVP with the time step $\delta t = 10.0$ ns.
To deal with the entanglement  growth during the time evolution, we use an adaptive bond dimension $\chi_\sigma$ with a maximal default value of $\chi^{\rm max}$ = 400 that may be changed according to the requirements of the given problem. The actual value of $\chi_\sigma$ at each $A[\sigma]$ is controlled by the uniform truncation parameter $\epsilon$, such that  discarded singular values of the SDV decomposition $\sum_{i\in \text{discarded}}\lambda_i^2 < \epsilon$ . Our setup allows users to chose between three compression levels $\epsilon \in \{10^{-8}, 10^{-10}, 10^{-12}\}$. 

\end{document}